\documentclass[11pt,bezier]{article}
\usepackage{amssymb,amscd,amsmath}
\def\NP#1#2{ Nucl.Phys. B#1 (#2)}
\def\PL#1#2{ Phys.Lett. B#1 (#2)}
\def\CMP#1#2{ Commun.Math.Phys. #1 (#2)}
\def\MPL#1#2{ Mod.Phys.Lett. A#1 (#2)}
\def\PRL#1#2{ Phys.Rev.Lett. #1 (#2)}

\def\IJMP#1#2{ Int.J.Mod.Phys. A#1 (#2)}
\newcommand{\pd}{\partial}
\newcommand{\bz}{\bar z}

\newcommand{\ap}{\alpha_{+}}
\newcommand{\am}{\alpha_{-}}
\newcommand{\an}{\alpha_{0}}
\newcommand{\apm}{\alpha_{+}^{\text m}}
\newcommand{\amm}{\alpha_{-}^{\text m}}
\newcommand{\tr}{\text{tr}}
\newcommand{\G}{\Gamma}
\newcommand{\g}{\gamma}
\newcommand{\mg}{\bar\gamma}
\newcommand{\bn}{\bar n}
\newcommand{\bm}{\bar m}

\newcommand{\D}{\Delta^{(0)}}

\newcommand{\2}{\sqrt 2}
\newcommand{\h}{\frac{1}{2}}
\newcommand{\kpz}{\Delta^{\text{KPZ}}}
\newcommand{\gd}{\text{dim}}

\title{ On touching random surfaces, two-dimensional quantum gravity and 
non-critical string theory} 
\author{Oleg Andreev\thanks{On leave from Landau Institute for 
Theoretical Physics, Moscow.}\\ \\
Department of Physics and Astronomy,\\
University of Southern California, Los Angeles, CA 90089-0484}
\date{}
\begin{document}
\maketitle

\vspace{-8cm}
\begin{flushright}
hep-th/9710107 \\
USC-97/HEP-A1
\end{flushright}

\vspace{8cm}
\begin{abstract}
A set of physical operators which are responsible for touching 
interactions in the framework of $c<1$ unitary conformal matter coupled 
to 2D quantum gravity is found. As a special case the non-critical 
bosonic strings are considered. Some analogies with four dimensional 
quantum gravity are also discussed, e.g. creation-annihilation operators 
for baby universes, Coleman mechanism for the cosmological constant. \\
PACS number(s): 04.60.-m, 11.25.Hf, 11.25.Pm
\end{abstract}


\vspace{1cm}
\section{Introduction}
\renewcommand{\theequation}{1.\arabic{equation}}
\setcounter{equation}{0}

In the last decade there has been much progress in understanding string 
theory in two dimensions as well as 2D quantum gravity 
(see, e.g. \cite{L} and references therein). Of course, for most physical 
applications one needs to consider much more complicated models, however 
many principal issues in string theory and quantum gravity are still not 
understood, and the hope is that the two dimensional theory will serve as 
a useful toy model, in which some of these issues may be addressed. For 
instance, a renormalization group (RG) approach developed for matrix models 
by Br\'ezin and Zinn-Justin \cite{EB} can be used to formulate a large $N$ 
renormalization group in a new M(atrix) theory \cite{Dou}. Another example 
of this is topological fluctuations in spacetime that produce baby 
universes. They were intensively discussed in a framework of four 
dimensional quantum gravity in relation with a theory of the cosmological 
constant and loss of quantum coherence \cite{Col}. Recently, it was 
proposed by David \cite{D1} that such fluctuations could lead to a scenario 
for the so-called $c=1$ barrier in two dimensions \cite{KPZ}. The work 
discussed in this paper was influenced by David's paper.

David begins with the renormalization group analysis of matrix models, with 
a new coupling constant that governs the dynamics of touching surfaces, i.e. 
surfaces which are allowed to touch each other at isolated points. In 
matrix models trace-squared terms are responsible for touching. For example, 
the one-matrix model with such interaction is given by \cite{MMM}
\begin{equation}
Z=\int{\cal D}\Phi\, exp \biggl[-N\tr(\frac{\Phi^2}{2}-g\frac{\Phi^4}{4})
-\frac{x}{2}(\tr\frac{\Phi^2}{2})^2\biggr]\quad.
\end{equation}
It is known that the model is solvable. Its phase diagram looks like


\hspace{.5cm}
\unitlength=1.00mm
\linethickness{0.4pt}
\begin{picture}(125.00,129.00)
\put(35.00,85.00){\vector(0,1){35.00}}
\put(30.00,90.00){\vector(1,0){85.00}}
\put(55.00,90.00){\line(0,-1){1.00}}
\put(75.00,90.00){\line(0,-1){1.00}}
\put(95.00,90.00){\line(0,-1){1.00}}
\put(35.00,100.00){\line(-1,0){1.00}}
\put(35.00,110.00){\line(-1,0){1.00}}
\put(120.00,90.00){\makebox(0,0)[cc]{$x$}}
\put(35.00,124.00){\makebox(0,0)[cc]{$g$}}
\put(29.00,110.00){\makebox(0,0)[cc]{0.1}}
\put(29.00,100.00){\makebox(0,0)[cc]{0.05}}
\put(55.00,85.00){\makebox(0,0)[cc]{0.25}}
\put(75.00,85.00){\makebox(0,0)[cc]{0.5}}
\put(95.00,85.00){\makebox(0,0)[cc]{0.75}}
\put(27.00,85.20){\makebox(0,0)[cc]{0}}
\bezier{200}(35.00,107.00)(53.00,91.00)(75.00,90.00)
\put(40.60,102.60){\circle*{1.00}}
\put(43.50,103.50){\makebox(0,0)[cc]{C}}
\put(49.00,108.00){\makebox(0,0)[cc]{gravity($c=0$)}}
\put(71.50,97.00){\makebox(0,0)[cc]{branched polymers}}
\put(40.60,85.00){\makebox(0,0)[cc]{$x_c$}}
\put(40.60,90.00){\line(0,-1){1.00}}

\end{picture}

\vspace{-8.5cm}
\begin{center}
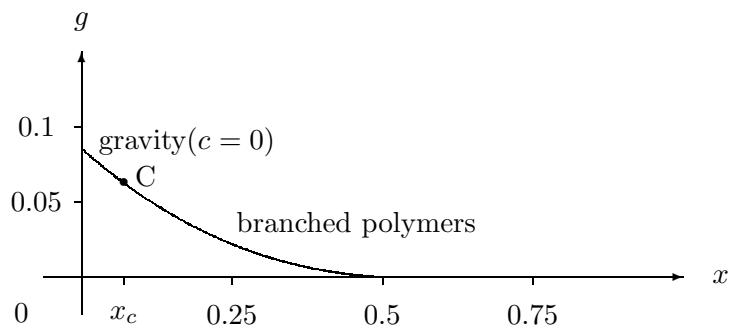
Fig.1. Phase diagram of the one-matrix model.
\end{center}

\vspace{.25cm}
The point C at $x=x_c$ corresponds to a critical behavior with the 
string exponent (string susceptibility) $\g=\frac{1}{3}$ \cite{MMM}. On 
the other hand critical lines $x<x_c$ and $x>x_c$ are characterized by 
the string exponents $\g=-\frac{1}{2}$ and $\g=\frac{1}{2}$, 
respectively. The first is described in terms of $c=0$ matter coupled 
to 2D gravity (pure gravity). As to the second, it is a branched 
polymer critical line. It should be noted that the multicritical 
point C appears due to fine-tuned touching interactions. At the same 
time touching is not very important for the pure gravity phase. In 
fact the above picture is valid for $c\leq 1$ models too.

It is well-known that the scaling for the critical lines with $\g<0$, 
associated with the conventional matrix 
models (no trace-squared terms), is described in terms of the Liouville 
effective action \cite{DDK}\footnote{Here and in the subsequent we 
restrict to the spherical topology. We also omit kinetic terms for 
matter in effective actions.}
\begin{equation}
S_{\text{ eff}}=\frac{1}{2\pi}\int d^2z \Bigl(\pd\phi\bar\pd\phi
-\frac{1}{4}Q\sqrt{\hat g}\hat R\phi+t_0\sqrt{\hat g}e^{\ap\phi}\Bigr)
\quad,
\end{equation}
where 
\begin{equation*}
\ap=\frac{1}{2\sqrt{3}}(\sqrt{1-c}-\sqrt{25-c})\quad ,\quad
Q=\sqrt{\frac{25-c}{3}}\quad.
\end{equation*}
$t_0$ is the renormalized cosmological constant. In the above we also 
assume that the unitary conformal matter has the central charge $c$. 
The string exponent is given by
\begin{equation}
\g=\frac{Q}{\ap}+2\quad .
\end{equation}

Klebanov et al.\cite{K} argued that the scaling for the multicritical 
points, associated with the modified matrix models, is also described 
in terms of the Liouville type action, but with a negatively dressed 
Liouville potential (cosmological term) namely,
\begin{equation}
\bar S_{\text{ eff}}=\frac{1}{2\pi}\int d^2z \Bigl(\pd\phi\bar\pd\phi-
\frac{1}{4}Q\sqrt{\hat g}\hat R\phi+\bar t_0\sqrt{\hat g}e^{\am\phi}\Bigr)
\quad,
\end{equation}
where
\begin{equation*}
\am=-\frac{1}{2\sqrt{3}}(\sqrt{1-c}+\sqrt{25-c})\quad.
\end{equation*}
It provides the string exponent
\begin{equation}
\mg=\frac{Q}{\am}+2\quad .
\end{equation}
Comparing to (1.3) one finds
\begin{equation}
\mg=\frac{\g}{\g-1}\quad ,
\end{equation}
which is in agreement with the matrix model results \cite{MMM}
\footnote{The same relation was also found in multiple spins on dynamical 
triangulations. The interested reader is referred to lectures of Ambj\o rn 
for details \cite{L}.}.

However the missing point of the continuum formulation sketched above is 
''touching'' operators, i.e. local operators which are responsible for the 
touching interactions. Our purpose is to show that the touching 
interactions can be reproduced in the continuum (Liouville) formulation 
too. At first sight, it seems naive that a network of touching surfaces 
is approximated by a surface with insertions of local operators as 
indicated in Fig.2. At the present time it is not


\unitlength=1mm
\linethickness{0.4pt}
\begin{picture}(113.00,72.24)
\put(40.00,55.00){\circle{14.00}}
\put(105.00,55.00){\circle{14.00}}
\put(80.00,55.00){\line(-5,3){5.00}}
\put(80.00,55.00){\line(-5,-3){5.00}}
\put(75.00,56.00){\line(0,2){2}}
\put(75.00,52.00){\line(0,2){2}}
\put(75.00,54.00){\line(-7,0){7}}
\put(68.00,56.00){\line(7,0){7}}
\put(49.00,62.00){\circle{8.49}}
\put(52.00,56.00){\circle{4.47}}
\put(29.00,60.00){\circle{10.00}}
\put(33.00,46.00){\circle{8.00}}
\put(24.00,54.00){\circle{6.00}}
\put(46.00,48.00){\circle{4.00}}
\put(109.50,49.50){\circle*{1.00}}
\put(110.00,60.00){\circle*{1.00}}
\put(98.50,57.50){\circle*{1.00}}
\put(100.00,50.00){\circle*{1.00}}
\put(114.00,60.00){\makebox(0,0)[cc]{${\text T}_2$}}
\put(113.00,49.00){\makebox(0,0)[cc]{${\text T}_3$}}
\put(95.10,57.00){\makebox(0,0)[cc]{${\text T}_1$}}
\put(99.00,47.00){\makebox(0,0)[rc]{${\text T}_4$}}
\end{picture}

\vspace{-4cm}
\begin{center}
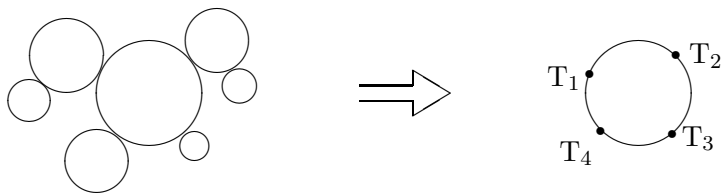
Fig.2. Approximation of a network of touching surfaces by a single 
surface with insertions of local operators ${\text T}_i$.
\end{center}

\vspace{.25cm}
\noindent  known whether the situation may be taken under control. Good 
motivations for this are the reproduction of the string exponents via 
Liouville and the rather special structure of surfaces when they touch each 
other at isolated points, i.e locally. So we are bound to learn something 
if we succeed.

Before continuing our discussion of the touching operators, we will make a 
detour and recall some basic results on 2D gravity coupled to $c\leq 1$ 
matter. 

First let us summarize notations for a matter sector. It is convenient to 
bosonize it as
\begin{equation}
S_{\text m}=\frac{1}{2\pi}\int d^2z \Bigl(\pd X\bar\pd X
+i\frac{1}{2}\an\sqrt{\hat g}\hat RX\Bigr)\quad,
\end{equation}
where $\an=\sqrt{\frac{1-c}{12}}$.
\noindent In this language the primary field of the conformal 
dimension $\D$ is represented as the exponent of the free field 
$X(z,\bz)$
\begin{equation}
V_{\alpha}(z,\bz)=e^{i\alpha X(z,\bz)}\quad,
\end{equation}
where $\D_{\alpha}=\frac{1}{2}\alpha(\alpha-2\an)$.

Dotsenko-Fateev models \cite{DF} arise at
\begin{equation}
\alpha=\alpha_{n.m}=\frac{1-n}{2}\amm+\frac{1-m}{2}\apm\quad,
\end{equation}
with integers $n,\,m$ and
\begin{equation}
\alpha_{\pm}^{\text m}=\frac{1}{2\sqrt 3}(\sqrt{1-c}\pm\sqrt{25-c})\quad.
\end{equation}
The corresponding primary fields are given by
\begin{equation}
V_{n.m}(z,\bz)=e^{i\alpha_{n.m} X(z,\bz)}\quad.
\end{equation}
Their conformal dimensions are written as
\begin{equation}
\D_{n.m}=\frac{1}{8}\biggl[(n\amm+m\apm)^2-(\amm+\apm)^2\biggr]\quad.
\end{equation}

Minimal models \cite{BPZ} are defined by $(\apm)^2=\frac{2q}{p}$, with 
the coprime integers $q$ and $p$. These models are very special because 
of the basic grid of the primary fields
\begin{equation*}
1\leq n\leq q-1 \quad,\quad\quad 1\leq m\leq p-1\quad.
\end{equation*}
Moreover, for conformal theories with $c<1$ there is a famous result of 
Friedan, Qiu and Shenker that the only unitary conformal theories with 
$c<1$ are the unitary series of the minimal models \cite{FQS}. They 
correspond to $q=p+1$ and have the central charge 
$c=1-\frac{6}{p(p+1)}\,$ with $p=2,3,\dots$.

Physical states in 2D gravity coupled to $c\leq 1$ matter were studied 
in the framework of the BRST quantization \cite{BRST}. There an 
important role is played by the BRST operator
\begin{equation*}
Q_{\text{BRST}}=\oint dz\,c(z)\Bigl(T_{\text{m}}(z)+T_{\text{L}}(z)+\h
T_{\text{gh}}(z)\Bigr)\quad,
\end{equation*}
where $T_{\text{m}}(z),\,\,T_{\text{L}}(z),\,\,T_{\text{gh}}(z)$ are the 
stress energy tensors for matter, Liouville and ghost sectors, 
respectively. The physical states (operators $\cal O$) are defined as 
the cohomology classes of this BRST operator. In this work we will 
mainly focus on the physical operators without ghost excitations, i.e. 
the tachyon and discrete states \cite{DS}. It is convenient to use a 
representation for such states when a matter sector is bosonized in a 
way as we sketched earlier. The tachyon type states are given by
\begin{align}
{\cal T}_{n.m}^{\pm}&=\int d^2z\,V_{n.m}(z,\bz)\,e^{\beta^{\pm}
(\D_{n.m})\phi(z,\bz)}\quad,\\
\beta^{\pm}(\D_{n.m})& =\frac{1}{2\sqrt{3}}(\pm\sqrt{1-c+24\D_{n.m}}
-\sqrt{25-c})\quad .
\end{align}
Since in the case of interest $\D_{n+p+1.m+p}=\D_{n.m}\,\,$, $n$ is 
restricted to a range $1\leq n\leq p+1\,$. Thus, one has the matter 
primaries $V_{n.m}$ not only inside the basic grid but also outside it. 
The discrete states appear as the border case operators $n=p+1$ or 
$m=0 \mod p$ (see e.g. \cite{D} and Appendix B). Notice that there are 
two independent Liouville exponents $\beta^{\pm}$, corresponding to two 
choices of dressing. From this point of view the scalings for the critical 
lines with $\g<0$ and multicritical points are described by the effective 
actions with the positively and negatively dressed Liouville potentials, 
respectively.

The outline of the paper is as follows. In sections 2.1 and 2.2 we describe 
touching interactions in the continuum. We not only reproduce the known 
matrix model results but find rather amusing new ones. Moreover, analogies 
with four dimensional quantum gravity appear. Section 3 will present the 
conclusions and directions for future work. In the appendices we give some 
technical details which are relevant for our discussion of the touching 
operators.


\section{Touching interactions in the continuum}
\renewcommand{\theequation}{2.\arabic{equation}}
\setcounter{equation}{0}

\subsection{$c<1$ models}
Let us now show how touching interactions appear in the continuum 
formulation. To do this, it is useful to begin with a geometrical 
analysis.

{\it 2.1.1. Geometry.} First of all, we turn to a geometrical 
interpretation of operators contained in the effective actions (1.2) 
and (1.4). It is well known that 
${\cal P}\equiv {\cal T}_{1.1}^{\pm}=\int d^2z\sqrt{\hat g}
e^{\alpha_{\pm}\phi}$ are called the puncture operators. A motivation 
for this is that an insertion of such operator into the path integral 
fixes a point on a Riemann surface. Such fixing corresponds to what in 
the theory of Riemann surfaces is called a puncture (see, e.g. \cite{W1} and 
references therein). This can be formulated in terms of the 
partition functions. Regarding $Z=\langle\,1\,\rangle$ as the partition 
function of an original surface, the partition function for the 
punctured surface is $Z_{\text {punc}}=\langle\,{\cal P}\,\rangle\,$. 
Note that this definition of the puncture operator differs from the 
one used in \cite{S} namely, $\int d^2z\sqrt{\hat g}\,V e^{-\frac{Q
}{2}\phi}$. They can only coincide at $c=1$ which is special because 
$\ap=\am=-\frac{Q}{2}=-\sqrt{2}$. 

Let us now look more specifically at touching interactions. 
Heuristically, the idea is that a network of touching spheres includes 
both the main surface (parent) as well as the pinched spheres attached to the 
parent (see Fig.2). It is well known that a surface attached to the 
parent by a wormhole (tiny neck) is called a baby universe \cite{Col}. 
However, in the context of two dimensional gravity the notion is 
simplified. A sphere attached to the parent is usually called a baby 
universe \cite{L}. In our case we also have pinched spheres attached to 
the parent. After this is understood, it immediately 
comes to mind to introduce a new notion. By analogy with the baby 
universe, we define a k-branched baby universe as the (k-1)-pinched 
sphere attached to the parent by a tiny neck\footnote{A motivation 
for such name is the structure of the attached surface. It allows to 
label the baby universes by the integer number $k$.}. Here we identify 
the standard baby universe with the 1-branched baby universe.

The distribution of the baby universes on a surface was analyzed 
in \cite{J} via dynamical triangulations. It was shown that the average 
number of minimum neck baby universes (whose neck thickness is of order 
of the ultraviolet cutoff) of area $B$ on a closed genus $g$ surface 
of area $A$ scales as 
\begin{equation}
N_A(B)\,\propto\,A^{3-\g(g)}(A-B)^{\g(g)-2}B^{\g-2}\quad,
\end{equation}
where $\g(g)=\g(1-g)+2g$.

We want now to repeat the analysis of ref.\cite{J} in order to find the 
average number of the minimum neck k-branched baby universes of area 
$B$ on a closed genus $g$ surface of area $A$. Note that the derivation 
is sufficiently generic, so one can apply it for both the critical 
lines and multicritical points (conventional and modified matrix 
models). We claim that
\begin{equation}
N_A(k,B)\,\propto\,A^{3-\G(g)}(A-B)^{\G(g)-2}B^{k\G-2}\quad,
\end{equation}
where $\G=\g,\mg$ and $N_A(1,B)\equiv N_A(B)$. The only fact needed to 
get (2.2) is that the partition function for the k-pinched sphere of area
 $A$ scales as $Z_k(A)\propto A^{(k+1)\G-3}$. It can be found 
repeatedly, reducing to the 1-pinched sphere via a sewing procedure. 
In the last case it is simply obtained by sewing two spheres with 
punctures. 

It follows from the statement (2.2) that the average number of the 
k-branched baby universes on the surface should scale as
\begin{equation}
N_A(k)=\int dB\,N_A(k,B)\,\propto\,A^{k\G}\quad.
\end{equation}

Suppose that the k-branched baby universes can be reproduced by a local 
operator. This means that its normalized one-point correlation function 
should scale as 
\begin{equation}
\langle\langle \,a_k^{\dagger}\,\rangle\rangle_A=
\frac{\langle\,a_k^{\dagger}\,\rangle_A}{\langle\, 1\,\rangle_A}
\,\propto\,A^{k\G}\quad.
\end{equation}
Here the symbol $\langle\,\,\,\,\,\rangle_A$ denotes the correlation 
functions computed using the actions (1.2) and (1.4) at fixed area $A$. 
$a_k^{\dagger}={\cal A}_k^{\dagger}\,,\,\bar{\cal A}_k^{\dagger}$, where 
${\cal A}\,,\,\bar{\cal A}$ correspond to the conventional and modified 
matrix models, respectively.
\newline On the other hand, this implies \cite{KPZ}
\begin{equation}
\langle\langle\,a_k^{\dagger}\,\rangle\rangle_A
\,\propto\,A^{1-\kpz_k}\quad.
\end{equation}
As a result, one finds that the KPZ scaling dimension of $a_k^{\dagger}$ 
is given by
\begin{equation}
\kpz_k=1-k\G\quad.
\end{equation}

It seems natural from physical point of view to call the 
$a_k^{\dagger}$'s as the creation operators as it was done in four 
dimensions \cite{Col}. Then, it immediately comes to mind 
to define the annihilation operators. A possible way to do this is to 
make use of two-point correlation functions. Let $a_k$ be the 
annihilation operators. Then two-point functions obey
\begin{equation}
\langle\langle\,\,a_k^{\dagger}a_k\,\,\rangle\rangle_A
\,\propto\, O(1)\quad.
\end{equation}
This allows one to find the KPZ scaling dimension of the operator $a_k$. It 
is given by
\begin{equation}
\kpz_k=1+k\G\quad.
\end{equation}
It should be stressed that a difference from the four dimensional case is 
that we define the $a_k$'s via a scalar product and 
not the standard commutation relations.

Of course, the annihilation operators can be defined by a geometrical 
analysis too. Let us give an example. Consider the case where a surface is 
the 1-pinched sphere; more complicated cases can be treated by a similar 
way. The number of the 1-pinched spheres of area $A$ scales as $A^{2\G -3}$. 
On the other hand, the number of degenerate 1-pinched spheres of the same 
area scales as $A^{\G -3}$. The latter assumes that the 1-pinched sphere 
degenerates into the sphere. It is clear because the baby universe 
vanishes. From the above statements, it follows that the average number of the 
degenerate 1-pinched spheres scales as $A^{\G -3}/A^{2\G -3}=A^{\G}$. This 
means that the normalized one-point function of the annihilation operator 
$a_1$ should scales as $A^{\G}$. As a result, we recover its KPZ scaling 
dimension $\kpz_1=1+\G$. 

{\it 2.1.2. Detailed examination of operators}. Let local operators which 
are responsible for the k-branched baby universes belong to the physical 
operators of 2D gravity coupled to conformal matter. It is known that 
such operators are characterized by the KPZ scaling 
dimension \cite{KPZ}. In the conformal gauge this dimension is 
completely defined by the $\phi$ zero mode \cite{DDK}. So, for the 
operator ${\cal O}_k$ with the Liouville
 exponent $\beta_k$, ${\cal O}_k\,\propto\,e^{\beta_k\phi_0}$, the KPZ 
scaling dimension is given by $\kpz_k=1-\frac{\beta_k}{\alpha}\,$, if 
the Liouville potential is $e^{\alpha\phi}$. In above $\phi_0$ is the 
zero mode of $\phi$.

For the critical lines with $\g<0$ (conventional matrix models) we 
use the above statements as well as equations (2.6), (2.8) in order 
to find the Liouville exponents of the creation and annihilation 
operators
\begin{equation}
{\cal A}_k^{\dagger}\,\propto\,e^{k(\ap-\am)\phi_0}
\quad,\quad\quad
{\cal A}_k\,\propto\,e^{k(\am-\ap)\phi_0}\quad.
\end{equation}

For the multicritical points (modified matrix models) similar 
calculations lead to
\begin{equation}
\bar {\cal A}_k^{\dagger}\,\propto\,e^{k(\am-\ap)\phi_0}
\quad,\quad\quad
\bar {\cal A}_k\,\propto\,e^{k(\ap-\am)\phi_0}\quad.
\end{equation}
It is interesting to note that all exponents vanish at $c=1$ that 
leads to $\kpz\vert_{c=1}=1$.

Let us now consider the partition function taking into account 
contributions from the branched baby universes. It is known that such 
configurations are present in the path integral over metrics. They 
correspond to singular world-sheet metrics. The partition function is 
given by
\begin{equation}
Z_{\text{pinched}}=\sum_{k=0}^{+\infty}{\it w}_k\,Z_k\quad,
\end{equation}
where $Z_k$ is a contribution of the k-pinched sphere. ${\it w}_k$ is a 
weight factor of each contribution. Suppose that $Z_{\text{pinched}}$ 
is described by the actions (1.2) and (1.4) perturbed by the creation 
and annihilation operators\footnote{In fact, the sums are finite 
(see (2.20) below).} 
\begin{align}
S_{\text{ eff}}^{\prime}&=S_{\text{ eff}}+
\sum_{k=1}^{+\infty}t_k\,{\cal A}_k+t^{\dagger}_k\,{\cal A}^{\dagger}_k
\quad ,\\
\bar S_{\text{ eff}}^{\prime}&=\bar S_{\text{ eff}}+
\sum_{k=1}^{+\infty} \bar t_k\,\bar{\cal A}_k
+\bar t^{\,\dagger}_k\,\bar{\cal A}^{\dagger}_k\quad.
\end{align}
Under this assumption, the gravitational dimensions of the coupling 
constants obey\footnote{This dimension is equal to $1-\kpz\,$.}
\begin{equation*}
\gd\, t_k>0\quad,\qquad
\gd\, t^{\dagger}_k<0\quad,\qquad
\gd\,\bar t_k <0\quad,\qquad
\gd\,\bar t^{\,\dagger}_k>0\quad,
\end{equation*}
from which it follows that the actions (2.12)-(2.13) are not 
renormalizable. However, if we define the theory as
\begin{align}
S_{\text{ eff}}^{\prime}&=S_{\text{ eff}}+
\sum_{k=1}^{+\infty}t_k\,{\cal A}_k
\quad ,\\
\bar S_{\text{ eff}}^{\prime}&=\bar S_{\text{ eff}}+
\sum_{k=1}^{+\infty} \bar t^{\,\dagger}_k\,\bar{\cal A}^{\dagger}_k
\quad,
\end{align}
then the actions are renormalizable. In other words, 
${\cal A}_k^{\dagger}$ and $\bar{\cal A}_k$ are the irrelevant operators 
that disappear in the IR limit. Such treating the actions 
leads us to a conclusion that the baby universes can be neglected for the 
critical lines with $\g <0$ but they are relevant for the 
multicritical points. This fact has been noted previously in the framework 
of the RG approach to matrix models \cite{D1}.

It is interesting to note that all couplings ($t^{\,\dagger}_k\,,\,\,
t_k\,,\,\,\bar t^{\,\dagger}_k\,,\,\,\bar t_k$) automatically become 
marginal at $c=1$. This is in accord with the conjecture on their 
role in the $c=1$ barrier. 

Finally, let us discuss a relation with the result of David \cite{D1}. 
To do this we must remember the definition of the scaling dimension 
in the framework of the renormalization group approach \cite{EB}. It 
is defined by
\begin{equation}
\Delta_x^{\text{RG}}=\frac{2\beta(\D_x)}{Q}\quad ,
\end{equation}
where $Q$ and $\beta(\D_x)$ are the background charge and Liouville exponent.

Combining this with (2.9) and (2.10), we learn that for the operators 
${\cal A}^{\dagger}_1\,,\,\, \bar{\cal A}^{\dagger}_1$ the scaling 
dimensions are simply
\begin{equation}
\Delta_1^{\text{RG}}=2\sqrt{\frac{1-c}{25-c}}\quad,\quad
\quad\bar\Delta_1^{\text{RG}}=-2\sqrt{\frac{1-c}{25-c}}\quad,
\end{equation}
which are the formulae derived in ref.\cite{D1}.

{\it 2.1.3. The examination, revisited.} Up to now our discussion has not
 been sensitive to a detailed structure of the physical operators ${\cal 
O}_k$. Suppose now that the touching operators are the tachyon type 
operators ${\cal T}_{n.m}^{\pm}$, i.e. they are given by the exponents of 
the free fields as in (1.13). Intuitively, this comes about because these 
operators are somewhat descendant from the puncture operators 
${\cal T}_{1.1}^{\pm}$. Indeed, the dimensions of the operators 
${\cal A}^{\dagger}_1\,,\,\,\bar{\cal A}^{\dagger}_1$ were obtained 
from the dimensions of ${\cal T}_{1.1}^{\pm}$ via 
a sewing procedure \cite{D1}. On the other hand the tachyon operators 
are the simplest physical operators in the theory and, moreover, they 
are moduli of the theory, so it is natural to start by looking for the 
touching operators among them. We will fill this gap in our 
determination of the touching operators in Appendix A.

Accepting the above assumption, an interesting conclusion which we can 
draw is that ${\cal A}_k^{\dagger}=\bar {\cal A}_k$ and 
${\cal A}_k=\bar {\cal A}_k^{\dagger}$. Indeed if the Liouville 
exponents are given by (2.9)-(2.10), then it follows from (1.13) that 
for the creation and annihilation operators, we get
\begin{gather}
{\cal A}_k^{\dagger}=\bar {\cal A}_k={\cal T}^+_{2k-1.2k+1}=
\int d^2z\,V_{2k-1.2k+1}(z,\bz)\,e^{k(\ap-\am)\phi(z,\bz)}\quad,\\
{\cal A}_k=\bar {\cal A}_k^{\dagger}={\cal T}^+_{2k+1.2k-1}=
\int d^2z\,V_{2k+1.2k-1}(z,\bz)\,e^{k(\am-\ap)\phi(z,\bz)}\quad.
\end{gather}
In particular, the operators introduced by David are simply 
${\cal T}^+_{1.3}$ and ${\cal T}^+_{3.1}\,$. 

It is interesting to note that the theory has the finite number of the 
creation-annihilation operators for the branched baby universes that 
means finite sums in (2.12)-(2.13). In fact, since in (1.13) $n$ belongs 
to the range $1\leq n\leq p+1$, the largest value of $k$ is given 
by\footnote{$[a]$ means the integer part of $a$.}
\begin{equation}
\text{max}\,k=
\begin{cases}
\bigl[\frac{p}{2}\bigr]+1& 
\text{for ${\cal A}_k^{\dagger}$ and $\bar {\cal A}_k$}\,\,,\\
\bigl[\frac{p}{2}\bigr]&
\text{for ${\cal A}_k$ and $\bar {\cal A}_k^{\dagger}$}\,\,.
\end{cases}
\end{equation}
It is clear that it is dependent of the matter central charge. In other 
words, a shape of world-sheets is determined by matter living on them.

So the two phases (critical lines with $\g<0$ and multicritical points) 
differ not only by a branch of gravitational dressing for the puncture 
operators (cosmological terms), but also by different roles of the same 
operators: creation (annihilation) of the branched baby universes in one 
case and their annihilation (creation) in the other. 

In order to take the 
assumption that the touching operators are the tachyon type ones into 
account completely it is advantageous to go in a slightly different way. 
Instead of using the geometrical point of view, we will follow 
renormalization group arguments and look for perturbations which become 
marginal at $c=1$.

Let us perturb the continuum theory, so that the effective actions 
(1.2) and (1.4) become
\begin{align}
S_{\text{ eff}}^{\prime}&=S_{\text{ eff}}+\sum_{m=-\infty}^{+\infty}
\sum_{n=1}^{p+1} t_{n.m}\,{\cal T}_{n.m}\quad ,\\
\bar S_{\text{ eff}}^{\prime}&=\bar S_{\text{ eff}}
+\sum_{\bm=-\infty}^{+\infty}\sum_{\bn=1}^{p+1}\bar t_{\bn.\bm}\,
{\cal T}_{\bn.\bm}\quad .
\end{align}
Here $t_{n.m},\bar t_{\bn.\bm}$ are renormalized couplings. 
${\cal T}_{n.m},\,\,{\cal T}_{\bn.\bm}$ denote the tachyon type operators 
defined in (1.13). 

Since the transition occurs at $c=1$ the gravitational dimensions of 
the couplings obey
\begin{equation*}
\gd\,t_{n.m}\vert_{c=1}=\gd\,\bar t_{\bn.\bm}\vert_{c=1}=0\quad.
\end{equation*}
These conditions are equivalent to 
\begin{equation}
\beta^{\pm}(\D_{n.m})\vert_{c=1}=\sqrt{\frac{1}{2p(p+1)}}\Bigl(
\pm\vert np-m(p+1)\vert-2p-1\Bigr)\vert_{p=\infty}=0\quad.
\end{equation}
There are two solutions of equation $\beta^+(\D_{n.m})\vert_{c=1}=0$ 
in the range $1\leq n\leq p+1$ namely, 
\begin{equation}
n=m\pm 2\quad,
\end{equation}
while equation $\beta^-(\D_{n.m})\vert_{c=1}=0$ has no solutions in this 
range. By substituting (2.24) into $\beta^+(\D_{n.m})$, we easily find 
the Liouville exponents
\begin{equation}
\beta^+(\D_{n.n\pm 2})=\frac{1\pm n}{2}(\ap-\am)\quad,
\end{equation}
where
\begin{equation*}
\ap=-\sqrt{\frac{2p}{p+1}}\quad,\quad \quad \am=-\sqrt{\frac{2(p+1)}{p}}
\quad.
\end{equation*}

The main new novelty of the above calculation is an appearance of 
operators with $n\,$ even. As we have seen the operators with $n$ odd 
are arisen by the idea on a role of the touching interactions in the 
$c=1$ barrier. From the geometrical point of view they correspond to the 
creation-annihilation operators for the branched baby universes. Now 
we would like to complement the discussion by including the operators 
${\cal T}^+_{n.n\pm 2}\,$ with even $n$ \footnote{It should be noted 
that $n=-m=1$ is special because 
$\beta^+(\D_{1.-1})\equiv 0$. As a result, the matter field is given by 
screening operators.}. In general, this issue is not completely 
understood. Here we can only speculate. The idea is heuristically 
that both string
 exponents of surfaces with boundaries and gravitational scaling 
dimensions of the operators for $n$ even have $\G /2$ as a unit 
of ''measurement''. It is natural therefore to relate these 
operators with holes on a surface. To illustrate this, consider 
a geometry in which a surface is made by 
pinching the hemisphere at a point on a boundary, as shown in Fig.3 
on the left. Such surface is reproduced by gluing the sphere to a point on 
the boundary of the hemisphere. It is easy to find the area dependence of 
the partition function for this case. It is given by 
$Z_{\h}(A)\propto A^{\frac{3}{2}\mg -3}$. On the other hand 
this scaling is recovered by inserting the operator ${\cal T}^+_{2.0}$ 
into the path integral for the sphere, as in Fig.3 on the right. This 
stimulates one to introduce a notion of a banged baby universe as the 
hemisphere attached to the parent by a point on the boundary and interpret 
${\cal T}^+_{2.0}$ as the creation operator for the banged baby universe. 
Since we restrict to the spherical topology, we leave the detailed analysis 
of these operators for future study. 

\vspace{.5cm}

\unitlength=1mm
\linethickness{0.4pt}
\begin{picture}(113.00,105.24)
\put(33.00,95.00){\circle{14.00}}
\put(47.00,95.00){\circle{14.00}}
\put(42.70,95.00){\circle*{8.00}}
\put(102.00,95.00){\circle{14.00}}
\put(80.00,95.00){\line(-5,3){5.00}}
\put(80.00,95.00){\line(-5,-3){5.00}}
\put(75.00,96.00){\line(0,2){2}}
\put(75.00,92.00){\line(0,2){2}}
\put(75.00,94.00){\line(-7,0){7}}
\put(68.00,96.00){\line(7,0){7}}
\put(109.15,95.00){\circle*{1.00}}
\put(116.00,95.00){\makebox(0,0)[cc]{${\cal T}^+_{2.0}$}}
\end{picture}

\vspace{-8.6cm}
\begin{center}
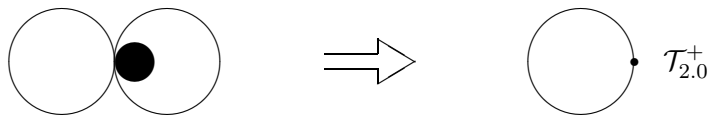
Fig.3. Approximation of the pinched hemisphere by the sphere with an
insertion of the local operator ${\cal T}^+_{2.0}$.
\end{center}

It is also not difficult to recognize the discrete state in 
${\cal T}^+_{2.0}$ \cite{DS}. This can be done 
using a linear map
\footnote{In fact, the map defined in (2.26) is a Lorentz boost in a two 
dimensional Minkowski space with coordinates $(X,\phi)$. We refer to 
\cite{L,P} for more details.}
\begin{equation}
X=\frac{Q}{2\2}{\mathbf{X}}-\frac{i\an}{\2}{\boldsymbol{\phi}}
\quad,\quad\quad
\phi=\frac{i\an}{\2}{\mathbf{X}}+\frac{Q}{2\2}{\boldsymbol{\phi}}\quad.
\end{equation}
Under this map one gets an effective $c=1$ matter dressed by gravity. 
In terms of the new variables the operator ${\cal T}^+_{2.0}$ becomes
\begin{equation}
{\cal T}^+_{2.0}=\int d^2z\,e^{i\2{\mathbf{X}}(z,\bz)}\quad.
\end{equation}
The holomorphic (anti-holomorphic) part of the integrand in (2.27) is 
the highest weight state of a spin-1 $su(2)$ multiplet. 

At this point, it is necessary to make a remark. One of the important  
statements about the discrete states was the following notice by 
Polyakov \cite{P}. The discrete states correspond to 
the contributions of singular world-sheet metrics, pinched spheres in 
the models under discussion, in the path integral over metrics. From our 
discussion of this issue, we have seen that there is, however, an 
important new feature that we must now clarify. We claim that for the 
unitary $c<1$ models in addition to the conventional discrete states there 
are a set of states ${\cal T}^+_{n.n\pm 2}\,$ which are also relevant. 
Moreover they are dominant. From the algebraic point of view the latter 
correspond to fractional values of the $su(2)$ spin.

{\it 2.1.4. Consequences.} Now we can easily read off some interesting 
conclusions. One of the first important observations is the following 
observation about a structure of the partition function 
$Z_{\text{pinched}}\,$. According to (2.20) there are no creation 
operators for the branched baby universes with $k$ larger than 
$\text{max}\,k\,$. It means that higher pinched spheres are obtained by 
attaching two or more creation operators to the parent. The 
effective action underlying such picture is given by
\begin{equation}
\bar S_{\text{ eff}}^{\prime}=\bar S_{\text{ eff}}+
\sum_{k=1}^{[\frac{p}{2}]} \bar t^{\,\dagger}_k\,
\bar{\cal A}^{\dagger}_k\quad\,\,\,.
\end{equation}
It should be stressed that this restriction is completely due to 
unitarity of the $c<1$ matter.

Next let us go on to look more carefully at the cases of interest. For the 
critical lines with $\g<0$ we find 
\begin{equation}
Z_{k+1}/Z_k\rightarrow 0\quad\text{under}\,\,t_0\rightarrow 0\quad,
\end{equation}
where $Z_k=\langle {\cal A}^{\dagger}_k\rangle\,$. So the leading 
contribution to $Z_{\text{pinched}}$ comes from the sphere, the next - from 
the pinched sphere etc. We belive that this fact allows
 one to interpret this phase  as the weak coupling regime for the 
touching interactions. This time they can contribute to subleading 
orders only. Formally the most relevant operator is 
${\cal A}^{\dagger}_1$. This is also in harmony with the idea of David 
that one should be able to catch effects of touching in this phase via 
this operator.

Now let us turn to the multicritical points. In contrast to the previous 
case $\bar{\cal A}^{\dagger}_k$ with $k=[\frac{p}{2}]$ is proven to be 
the most relevant operator in (2.28)\footnote{It is 
straightforward to get this result in the framework of the RG approach 
\cite{EB}. Looking at the scaling dimensions $\Delta_k^{\text{RG}}=
-\frac{k+1}{2p+1}\,$, we see that $\bar t^{\,\dagger}_{[\frac{p}{2}]}$ is 
the most relevant. However it is less relevant in comparison with the 
cosmological constant $\bar t_0$.}. From the geometrical point of view 
it means that the most branched baby 
universes are dominant. As a result, the expansion of 
$Z_{\text{pinched}}$ in powers of $\bar t^{\,\dagger}_k$ as it may follow 
from the action (2.28) is not valid. So we no longer have
the weak coupling regime for the touching interactions. Instead 
of this we interpret this phase as the strong coupling regime for the 
touching interactions. At this point, it is necessary to discuss a relation 
with the David scenario where the touching interactions were taken into 
account by the baby universes, i.e. $\bar{\cal A}^{\dagger}_1$. In our 
consideration of this issue, we have seen that the most relevant operator 
is $\bar{\cal A}^{\dagger}_{[\frac{p}{2}]}\,$. The latter means that the 
David picture is valid at least for the pure gravity ($p=2$) and Ising 
model ($p=3$). However, for $p\geq 4$ this can not be the hole story, 
for a reason that the branched baby universes come into the game and 
moreover, they are dominant.

Finally, let us note that the conclusion by Klebanov \cite{K} that the 
scaling limits of the conventional matrix models (critical lines with 
$\g<0$) and modified matrix models (multicritical points) differ due to the 
branches of gravitational dressing for the Liouville potential can be 
extended. According to our discussion, these scaling limits correspond to 
different phases of the touching interactions namely, weak and strong coupling 
regimes.

\subsection{Strings}
We now turn to the problem of shedding some light on touching interactions 
for $c=1$ models. It is well known that such models are non-critical 
bosonic strings or, equivalently, two dimensional critical 
strings \cite{L}. Thus, we will try to analyze effects of singular 
world-sheet metrics (pinched spheres) in the Polyakov path integral. In 
doing so, we will not follow the geometrical analysis of subsect. 2.1.1. 
Instead of this we look for the limit $p\rightarrow\infty\,$.

{\it 2.2.1. $p\rightarrow\infty$ limit}. One of the novelties that appears 
at $c=1$ is that the string exponents defined in (1.3) and (1.5) vanish. 
As a result, direct use of the geometrical point of view fails. Moreover, 
scaling violations for the phase associated with the conventional 
matrix models 
are also a serious obstacle on this way. In finding touching interactions 
for $c=1$ models it seems sensible to take as a starting point the model of 
sect. 2.1 for arbitrary $p$ then define the limit $p\rightarrow\infty\,$. 
To see what really happens, consider the effective actions. The Liouville
 exponents $\alpha_{\pm}$ will be $-\2\,$. One can imagine that the 
effective actions $S_{\text{ eff}}$ and $\bar S_{\text{ eff}}$ coincide but 
it is not true. As Polchinski pointed out \cite{Pol}, the Liouville 
potential for $S_{\text{ eff}}$ is given by $\phi e^{-\2\phi}$ that leads 
to the scaling violations. On the other hand, there are no scaling 
violations for the phase associated with the modified matrix models, so the 
potential for $\bar S_{\text{ eff}}$ is simply $e^{-\2\phi}\,$ \cite{K}. 
Thus, one has for the effective actions (1.2) and (1.4) at $c=1$
\begin{align}
S_{\text{ eff}}&=
\frac{1}{2\pi}\int d^2z \Bigl(\pd\phi\bar\pd\phi-\frac{1}{\2} 
\sqrt{\hat g}\hat R\phi+t_0\sqrt{\hat g}\phi e^{-\2\phi}\Bigr)
\quad ,\\
\bar S_{\text{ eff}}&=
\frac{1}{2\pi}\int d^2z \Bigl(\pd\phi\bar\pd\phi-\frac{1}{\2}\sqrt
{\hat g}\hat R\phi+\bar t_0\sqrt{\hat g}e^{-\2\phi}\Bigr)
\quad \,\,\,,
\end{align}
with the background charge $Q=2\2\,$.

Now we come to the analysis of the actions (2.12) and (2.13). Obviously, 
under the limit $p\rightarrow\infty$ these actions are given by
\begin{align}
S_{\text{ eff}}^{\prime}&=S_{\text{ eff}}+
\sum_{k=1}^{\infty}t_k\,{\cal A}_k+t^{\dagger}_k\,{\cal A}^{\dagger}_k
\quad ,\\
\bar S_{\text{ eff}}^{\prime}&=\bar S_{\text{ eff}}+
\sum_{k=1}^{\infty}\bar t_k\,\bar{\cal A}_k+
\bar t^{\,\dagger}_k\,\bar{\cal A}^{\dagger}_k\quad,
\end{align}
with the operators 
\begin{equation}
\bar {\cal A}_k={\cal A}^{\dagger}_k=
\int d^2z\,e^{-i\2 X(z,\bz)}\quad,\quad\quad
{\cal A}_k=\bar {\cal A}^{\dagger}_k=\int d^2z\,e^{i\2 X(z,\bz)}\quad.
\end{equation}
There is an interesting observation related with vanishing of the string 
exponents that the operators are independent of $k\,$. 
In other words, one can not distinguish the branched baby universes 
at $c=1\,$. Instead of this, there are collective potentials for 
the touching interactions with the following effective couplings:
\begin{equation*}
t=\sum_{k=1}^{\infty}t_k\quad,\qquad
t^{\dagger}=\sum_{k=1}^{\infty}t^{\dagger}_k\quad,\qquad
\bar t=\sum_{k=1}^{\infty} \bar t_k \quad,\qquad
\bar t^{\dagger}=\sum_{k=1}^{\infty} \bar t^{\,\dagger}_k\quad. 
\end{equation*}
The actions (2.32) and (2.33) are rewritten as
\begin{align}
S_{\text{ eff}}^{\prime}&=S_{\text{ eff}}+
t\int d^2z\,e^{i\2 X(z,\bz)}+
t^{\dagger}\int d^2z\,e^{-i\2 X(z,\bz)}
\quad ,\\
\bar S_{\text{ eff}}^{\prime}&=\bar S_{\text{ eff}}+
\bar t\int d^2z\,e^{-i\2 X(z,\bz)}+
\bar t^{\dagger}\int d^2z\,e^{i\2 X(z,\bz)}
\quad\,\,\,\,.
\end{align}
Thus, we have generalized the touching operators to $c=1$ models. At 
this point a few comments are in order:
\begin{enumerate}
\item[(i)] It is interesting to note that the holomorphic 
(anti-holomorphic) parts of the touching operators for the string models 
are non others than the screening operators of the $c=1$ conformal 
field theory (matter sector in the particular case at hand). It is well 
known that they represent the raising and lowering operators of the 
$su(2)$ algebra and generate the multiplets of the discrete states 
(see Appendix B for details). From this point of view our 
introduction of the annihilation operators seems plausible. However such 
operators do not lead to the standard Heisenberg algebra as it happens in a 
framework of four dimensional quantum gravity, but $su(2)\,$.
\item[(ii)] According to our discussion in subsect. 2.1.4, the weak and 
strong coupling regimes for the touching interactions are associated with 
the conventional and modified matrix models for $c<1\,$. At $c=1$ 
relations which are similar to (2.29) are not valid anymore. Instead of 
them, we have $Z_{k+1}/Z_k\sim 1$ that indicates the presence of a boundary 
between these 
phases. However this boundary looks singular because one does not get into
 the same theory under the $p\rightarrow\infty$ limit. 
\item[(iii)] If one makes use of a perturbation of the actions 
(2.35)-(2.36) according to which the creation and annihilation operators 
are involved with the same effective coupling constant a result will 
be the sine-Gordon model coupled to 2D gravity! Thus, the sine-Gordon model 
coupled to 2D gravity is an appropriate framework to take into account 
effects of singular world-sheet metrics in the Polyakov path integral for 
the non-critical strings. Unfortunately one knows very little about 
integrable models in the presence of quantum gravity. Some issues have 
been discussed in \cite{L,Mr,Bil}.
\end{enumerate}

{\it 2.2.2. The cosmological constant and touching interactions.} There is 
a serious problem in quantum gravity related with the vanishing of the 
cosmological constant. It is known several different proposals to solve it. 
One of them is based on the idea of uncontrollable emissions of tiny baby 
universes. It was intensively discussed in a framework of four dimensional 
case (see e.g. \cite{Col}).

Let us now try a two dimensional case. It is well known that the 
cosmological constant is being renormalized in a singular way 
as\footnote{In the literature on the $c=1$ models the cosmological term 
(puncture operator) is usually chosen as $e^{-\2\phi}$ which in our language 
corresponds to the multicritical points.}
\begin{equation}
\frac{\bar t_0}{\G [0]}=\bar {\mathbf t}_0 \quad,
\end{equation}
where $\G [x]$ is the gamma function. The origin of this multiplicative 
renormalization is of course the short distance divergences. In 
calculating of amplitudes one needs to perform multiple integrals. 
There are some prescription to do this. On of them is an analytic 
continuation. Shifting the exponents of the integrals one brings them 
into a standard Dotsenko-Fateev form. Next, the integrals are computed 
by an analytic continuation.

We are going to find the multiplicative renormalization of the touching 
couplings. In order to do this, we follow a similar procedure as it was 
used to derive (2.37). The calculation for this case, see the 
Appendix B, leads to the result:
\begin{equation}
\bar t\,\G [0]=\bar\mathbf t\quad,\qquad
\bar t^{\dagger}\,\G [0]=\bar\mathbf t^{\dagger}\quad.
\end{equation}
We see that the bare cosmological constant and touching couplings are 
renormalized in different ways namely, the cosmological constant goes 
to be ''zero'' but the touching interaction couplings go to 
''infinity''. Here an analogy with the four dimensional case 
appears again because such behavior reminds one of Coleman's 
idea that touching 
interactions (wormholes) have the effect of making the cosmological 
constant vanish \cite{Col}. Although it looks in many ways attractive, 
we have to stress its speculative character. It rests on the multiplicative 
renormalization argument only, so further work needed to prove it 
strictly.


\section{Conclusions and remarks}
\renewcommand{\theequation}{3.\arabic{equation}}
\setcounter{equation}{0}
First, let us say a few words about the results.

In this work we have found a set of the physical operators which are 
responsible for the touching interactions in the framework of $c<1$ 
unitary conformal matter coupled to 2D quantum gravity. It turned out 
that one can interpret the critical lines with $\g <0$ (conventional 
matrix models) and multicritical points (modified matrix models) as 
different phases for touching namely, the weak and strong coupling 
regimes. Next we defined the touching operators for the non-critical 
bosonic strings. It shows that if the creation and annihilation operators 
are involved with the same effective coupling constant. Then the sine-Gordon 
model coupled 
to 2D gravity is an appropriate framework to take into account effects 
of singular world-sheet metrics in the Polyakov path integral. Some 
analogies with the four dimensional case are also discussed, e.g.
the creation-annihilation operators for the baby universes and Coleman 
mechanism for the cosmological constant.

Let us conclude by mentioning a few open problems together with 
interesting features of the touching interactions in the continuum.
\begin{enumerate}
\item[(i)]Of course, the most important open problem is to understand 
the touching interactions in the critical strings or how to take into 
account effects of singular world-sheet metric in the Polyakov path 
integral. Unfortunately it is unknown in general how to realize this 
program. Our analysis of section 2 essentially rests on the Liouville 
mode $\phi$, so any attempt to use it for critical strings will fail.
\item[(ii)]In order to calculate the multiplicative renormalizations of 
the coupling constants we found special correlators of the discrete 
states. This seems strange because it is possible to find them directly 
from the action (2.36). However, by calculating correlators we solve one 
more problem which is formulated as the deformation of the OP algebra of 
the discrete states by the presence of non-vanishing cosmological and 
touching coupling constants. Although a special solution is 
known \cite{D,W} the problem is still open. Some progress in this direction 
has already done \cite{A}.
\item[(iii)] The operator ${\cal T}^+_{3.1}$ is special because it 
interpolates between matrix models \cite{DK}. In the simplest case it 
describes the flow from Ising ($p=3$) to pure gravity ($p=2$). We 
offer a qualitative physical interpretation of such transition based on 
our geometrical picture. First let us recall that a shape of world-sheets 
depends on the central charge of matter living on them namely, higher 
pinched world-sheets correspond to higher central charges. Next note that 
${\cal T}^+_{3.1}$ is nothing but the annihilation operator for the baby 
universes in the framework of the conventional matrix models, so it smooths 
a shape that leads to a proper reducing of the central charge. As a result, 
one has the flow from Ising to pure gravity. On the other 
hand it is the creation operator in the context of the modified matrix 
models, so it wrinkles a shape that increases the central charge. This time 
there is the flow from pure gravity to Ising. Of course, these conclusions 
are heuristic and further work needed to make them more rigorous.
\end{enumerate}


{\bf {Acknowledgments}}

I am indebted to I.Bars, J.-L.Gervais, V.Kazakov, R.Metsaev for useful 
discussions, and especially F.David for his comments on an earlier draft of 
the paper. I am also grateful to G. Lopes Cardoso for 
reading the manuscript. This research was supported in part by NRC 
grant GAC021197 and by Russian Basic Research Foundation under 
grant 96-02-16507.


\vspace{.3cm}
\appendix{{\bf Appendix A}}
\renewcommand{\theequation}{A.\arabic{equation}}
\setcounter{equation}{0}

\vspace{.3cm}
In discussing touching interactions, we assumed in subsect. 2.1.3 that 
local operators which are responsible for the branched baby universes are 
the tachyon type physical operators. In the present appendix, we will 
analyze some aspects of this story in somewhat more depth.

To begin with, we review some facts about the BRST formalism 
\cite{BRST,WZ}. The physical states are the cohomology classes of the 
BRST operator $Q_{\text{BRST}}$ whose explicit form is given in section 1. 
These classes are labeled by the ghost number $G$. The tachyon and 
discrete operators appear at ghost number two. So the operators (1.13) 
are rewritten as
\begin{equation}
{\cal T}_{n.m}^{\pm}(z,\bz)=c(z)\bar c(\bz)V_{n.m}(z,\bz)\,
e^{\beta^{\pm}(\D_{n.m})\phi(z,\bz)}\quad.
\end{equation}
Such class was intensively discussed in subsect. 2.1.3.

As for the new BRST classes, the first nontrivial example appears 
at $G=0\,$. These operators are denoted as 
${\cal O}_{j.m}\bar {\cal O}_{j.m}\,$. It is well known that the 
holomorphic (anti-holomorphic) operators ${\cal O}_{j.m}$ 
($\bar {\cal O}_{j.m}$) generate the chiral (anti-chiral) 
ground ring \cite{GR} namely, 
${\cal O}_{j_1.m_1}{\cal O}_{j_2.m_2}={\cal O}_{j_1+j_2.m_1+m_2}\,$. 
This allows one to determine the explicit form of an arbitrary operator from 
the first few which are given by\footnote{Note that 
${\cal O}_{0.0}\equiv 1\,$.}
\begin{align}
{\cal O}_{\h.\h}&=\Bigl( cb+\frac{i}{\2}\pd\mathbf{X}-\frac{1}{\2}\pd
\boldsymbol{\phi}\Bigr)
e^{(\frac{i}{\2}\mathbf{X}+\frac{1}{\2}\boldsymbol{\phi})}
\quad\,\,\,\,,\\
{\cal O}_{\h.-\h}&=\Bigl( cb-\frac{i}{\2}\pd\mathbf{X}-\frac{1}{\2}\pd
\boldsymbol{\phi}\Bigr)
e^{(-\frac{i}{\2}\mathbf{X}+\frac{1}{\2}\boldsymbol{\phi})}
\quad,
\end{align}
where $(\mathbf{X},\boldsymbol{\phi})$ refer to the effective $c=1$ 
matter dressed by gravity. In order to translate these operators into 
the $c<1$ theory, one can use a linear map
\begin{equation}
\mathbf{X}=\frac{Q}{2\2}X +\frac{i\an}{\2}\phi
\quad,\quad\quad
\boldsymbol{\phi}=-\frac{i\an}{\2}X+\frac{Q}{2\2}\phi
\quad ,
\end{equation}
which is inverse to (2.26). However, we do not need to do this. It is 
easy to understand that the operators 
${\cal O}_{j.m}\bar {\cal O}_{j.m}\,$ are not responsible for the 
branched baby universes. Indeed, they have nonzero Liouville exponents 
at $c=1$, so they can not be written as in (2.9)-(2.10).

Up to now we have discussed only a part of the BRST cohomology. Another 
part is recovered by the operator $a+\bar a$ \cite{WZ}, where
\begin{equation}
a=c(-i\an\pd X+\frac{Q}{2}\pd\phi)+2\pd c\quad.
\end{equation}
It is BRST invariant. So, applying $a+\bar a$ to ${\cal T}_{n.m}^{\pm}$ 
and ${\cal O}_{j.m}\bar {\cal O}_{j.m}$ one can form the new families 
of BRST invariant (physical) operators\footnote{To be precise, 
$a{\cal O}(0)=\oint_{{\cal C}_0}\frac{dz}{z}\,a(z){\cal O}(0)\,$; 
the contour ${\cal C}_0$ surrounds $0\,$.} 
$(a+\bar a){\cal T}_{n.m}^{\pm}$ with $G=3$ and 
$(a+\bar a){\cal O}_{j.m}\bar {\cal O}_{j.m}$ with $G=1\,$. Obviously, 
they have the same Liouville exponents as ${\cal T}_{n.m}^{\pm}$ and 
${\cal O}_{j.m}\bar {\cal O}_{j.m}\,$. Due to this reason the 
operators $(a+\bar a){\cal O}_{j.m}\bar {\cal O}_{j.m}$ are not 
appropriate for a role of the touching operators. As for the 
$(a+\bar a){\cal T}_{n.m}^{\pm}$'s, as their Liouville exponents are 
fitted to (2.9)-(2.10) at $n=m\pm 2\,$, they may be responsible for the 
branched baby universes. So there is a puzzle here. Before continuing 
our discussion of this puzzle, we wish to complete the review of the BRST 
cohomology classes.

Given a state with ghost number $G$ and Liouville exponent $\beta$, the 
two point function on the sphere defines a dual state with ghost number 
$6-G$ and Liouville exponent $-Q-\beta\,$\cite{WZ}. One immediately see 
that, with the Liouville exponents as defined in (1.14), 
$\beta^-(\D)=-Q-\beta^+(\D)\,$. So this definition provides a pairing 
between the positively and negatively dressed states. Note that $6$ comes 
from the ghost zero modes on the sphere while $-Q$ appears from the 
Liouville background charge\footnote{Notice that the dual states 
arising under factorization of correlation functions are not the ones 
defined via the two point functions on the sphere but states differing 
from them by $b_0-\bar b_0\,$. It leads to ghost number $5-G$.}. Using 
such procedure it is possible to find two new BRST cohomologies classes 
${\cal P}_{n.m}\bar {\cal P}_{n.m}$ and 
$(a+\bar a){\cal P}_{n.m}\bar {\cal P}_{n.m}$ at ghost number four and 
five, respectively \cite{WZ}. Since ${\cal P}_{n.m}\bar {\cal P}_{n.m}$ 
are dual to ${\cal T}_{n.m}^+$ it implies that they are the negatively 
dressed states with the Liouville exponents $\beta^-(\D_{n.m})$. At 
these values of the exponents, it is impossible to satisfy (2.9)-(2.10). 
This follows from the fact that $k(\ap-\am)\vert_{c=1}=0$ while 
$\beta^-(\D_{n.m})$ never vanishes at $c=1$ for $1\leq n\leq p+1\,$. 
So, the operators ${\cal P}_{n.m}\bar {\cal P}_{n.m}$ are not 
appropriate for the touching operators. For essentially this reason the 
operators $(a+\bar a){\cal P}_{n.m}\bar {\cal P}_{n.m}$ are also 
rejected. However, it is not the hole story about the BRST cohomology. 
Witten and Zwiebach found that there exist BRST invariant operators 
which can not be written as products of the holomorphic and 
anti-holomorphic operators \cite{WZ}. If ${\cal Y}_{n.m}^{\pm}$ denotes 
the holomorphic part of the operator ${\cal T}_{n.m}^{\pm}$ defined 
in (A.1), then the rest of the BRST cohomology is given by 
\begin{equation}
{\cal Y}_{n.m}^+\bar{\cal O}_{\bn.\bm}\,,\quad
{\cal O}_{n.m}\bar{\cal Y}_{\bn.\bm}^+\,,\quad
{\cal Y}_{n.m}^-\bar{\cal P}_{\bn.\bm}\,,\quad
{\cal P}_{n.m}\bar{\cal Y}_{\bn.\bm}^-\,
\end{equation}
and their products with $(a+\bar a)$. It is well known that in tensoring 
together holomorphic and anti-holomorphic operators (left and right 
moving states), one should restrict to operators of equal ''left'' and 
''right'' Liouville exponents. This allows one to reject these operators 
by the same arguments as it was done for 
${\cal O}_{j.m}\bar {\cal O}_{j.m}\,,\,\,{\cal Y}
_{n.m}^-\bar{\cal Y}_{n.m}^-\,,\,\,{\cal P}_{n.m}\bar {\cal P}_{n.m}$ and
 their products with the operator $(a+\bar a)$ in above. 

Summarizing, we have two classes of the BRST invariant operators which 
may formally be the touching operators namely, ${\cal T}_{n.n\pm 2}^+$ 
and $(a+\bar a){\cal T}_{n.n\pm 2}^+\,$. It remains to make our choice. 
Before doing it, let us discuss two points. 

First, let us recall what we want. Our goal is to describe a network of 
touching surfaces by a single surface (parent) with insertions of local 
operators. Moreover, we would like to have a field theory description, 
i.e. an effective action whose terms are responsible for pinched spheres 
attached to the parent. 

Next, let us turn to moduli. We recall that the moduli are operators that
 can be added to the action of the conformal field theory. In the 
particular case at hand they come from spin zero operators of ghost 
number two \cite{WZ}. For the operators ${\cal T}_{n.n\pm 2}^+(z,\bz)$ 
defined in (A.1) the corresponding moduli are 
$V_{n.n\pm 2}(z,\bz)\,e^{\beta^+(\D_{n.n\pm 2})\phi(z,\bz)}\,$, i.e. 
they are the integrands of the tachyon type operators (1.13)
! It is clear that this is precisely what we need. Thus, the touching 
operators are given by ${\cal T}_{n.n\pm 2}^+\,$.


\vspace{.3cm}
\appendix{{\bf Appendix B}}
\renewcommand{\theequation}{B.\arabic{equation}}
\setcounter{equation}{0}

\vspace{.3cm}
The purpose of this appendix is to compute the multiplicative 
renormalization of the touching couplings. It turns out that it is easy 
to find it by computing correlators of the discrete states of the $c=1$ 
models.

To begin with, let us remind how the discrete states appear in the 
theory. Taking the limit $p\rightarrow\infty$ one has for the matter 
sector (see (1.9)-(1.10))
\begin{equation}
\apm=-\amm=\2\quad,\quad\quad \alpha_{n.m}=\2 j\quad,\quad\quad
j\equiv\frac{n-m}{2}\quad.
\end{equation}
In addition the primaries (1.11) are rewritten as $V_{j.\pm j}(z,\bz)
=e^{\pm i\2 jX(z,\bz)}$. 

It is well known that the theory has ${\hat {su}}(2)\oplus 
{\hat {su}}(2)$ as the symmetry algebra. The holomorphic currents are
\begin{equation}
H^{\pm}(z)=e^{\pm i\2 X(z)}\quad,\quad\quad
H^0=\frac{i}{\2}\pd X(z)\quad.
\end{equation}
Obviously, their zero modes $H^a=\oint dz\,H^a(z)$ generate the su(2) 
algebra\footnote{We use the normalization 
$\oint_{{\cal C}_0}\frac{dz}{z}=1$ and omit $2(\pi)$ when it is 
irrelevant in the context of the present work.}. $H^{\pm}$ also 
play a role of the screening operators 
of the $c=1$ conformal field theory. 

It was realized for a long time ago \cite{G} that the primary fields form
 tensor products of $su(2)$ multiplets (holomorphic and anti-holomorphic)
\begin{gather}
V_{j.m}(z,\bz)={\cal N}_0(j.m)(H^-\bar H^-)^{j-m}V_{j.j}(z,\bz)
\quad,\\
{\cal N}_0(j.m)=\frac{(j+m)!}{(2j)!(j-m)!}\quad,\quad\quad
j=0,\,\h,\,1,\,\dots\quad,
\end{gather}
such that only $V_{j.\pm j}$ are the tachyon type primary fields defined 
in (1.11). As to the others, they are ''discrete primaries''.

Now, let us couple the $V_{j.m}$'s to gravity. It can be done directly, 
using the formulae (1.13)-(1.14). As a result, one gets
\begin{equation}
{\cal T}_{j.m}^{\pm}={\cal N}_1(j.m)
\int d^2z\,V_{j.m}(z,\bz)\,e^{\beta^{\pm}(j)
\phi(z,\bz)}\quad,\quad\quad
\beta^{\pm}(j)=\2(-1\pm j)\quad .
\end{equation}
Here the normalization factors ${\cal N}_1(j.m)=(2j)!(j+m)!(j-m)!$ are 
introduced to have the following OP algebra of the integrands
\begin{equation}
{\cal T}_{j_1.m_1}^+(z,\bz){\cal T}_{j_2.m_2}^+(0)=\frac{1}{\vert z\vert ^2}
(j_1m_2-j_2m_1){\cal T}_{j_1+j_2-1.m_1+m_2}^+(0)\quad,
\end{equation}
with vanishing value of the cosmological constant as well as touching 
couplings \cite{GR,KP}. 

In order to find the multiplicative renormalizations of couplings let 
us compute a few terms on the right hand side of (B.6) due to the 
presence of the non-vanishing cosmological and touching coupling 
constants\footnote{The deformation of this algebra only by 
the non-vanishing cosmological constant was found in \cite{D,W}.}. The 
coefficient at ${\cal T}_{j_3.-m_3}^{\pm}$ is given by
\begin{equation}
\langle {\cal T}_{j_1.m_1}^+(0){\cal T}_{j_2.m_2}^+(1)
\tilde  {\cal T}_{j_3.m_3}^-(\infty)\rangle\quad,
\end{equation}
with a conjugate operator defined as 
\begin{equation*}
\tilde  {\cal T}_{j.m}^-(z,\bz)=\tilde {\cal N}_1(j.m)
(H^+\bar H^+)^{j+m}\,V_{j.-j}(z,\bz)\,e^{-\2 (1+j)\phi(z,\bz)}
\quad,\quad \tilde {\cal N}_1(j.m)=[(2j)!(j+m)!]^{-2}\quad.
\end{equation*}
To find it, one can expand $e^{-\bar S_{\text{ eff}}^{\prime}}$
in powers of $\bar t_0\,,\,\,
\bar t\,,\,\,\bar t^{\dagger}$ and interpret the resulting terms as 
correlation functions in the free theory. 

As a warm up, let us reproduce the multiplicative renormalization of 
the cosmological constant. Following Dotsenko \cite{D}, 
set $m_1=j_1\,,\,\,
m_2=j_3-j_2\,,\,\,m_3=-j_3\,$. It is clear that the normalization 
factors don't lead to $\G [0]\,$, so we drop them. The contribution 
of the matter sector is given by
\begin{equation}
\G ^2[j_1+j_2-j_3+1]\prod_{i=1}^{j_1+j_2-j_3}
\frac{\G ^2[i]\,\G ^2[2j_3+i]}{\G ^2[2j_1+1-i]\,\G ^2[2j_2+1-i]}+
O(\bar t\,\bar t^{\dagger})\quad.
\end{equation}
It also doesn't lead to $\G [0]$ (at least in the leading order of 
$\bar t\,\bar t^{\dagger}$). On the other hand the Liouville sector contributes
\begin{equation}
\Bigl(\frac{\bar t_0}{\G [0]}\Bigr)^{j_1+j_2-j_3-1}
\prod_{i=1}^{j_1+j_2-j_3-1}
\frac{\G ^2[2j_1-i]\,\G ^2[2j_2-i]}{\G ^2[1+i]\,\G ^2[2j_3+1+i]}\quad.
\end{equation}
This expression shows that one has the multiplicative renormalization 
(2.37) for the cosmological constant. Note that such computation is 
an old story \cite{D}. The only novelty is contributions of the 
touching operators in (B.8). However, they can be neglected.

Now let us turn to the touching couplings. In contrast to the previous 
case set $m_1=j_1\,,\,\,m_2=j_2\,,\,\,m_3=-j_3$ and, moreover, 
$j_3=j_1+j_2-1\,$. The normalization factors don't give $\G [0]$ 
and we drop them again. The Liouville correlator is trivial. So, the 
only contribution is due to the matter sector. It is given by
\begin{equation}
\frac{\G ^2[2j_3+1]}{\G ^2[2j_1]\,\G ^2[2j_2]}
\sum_{k=0}^{\infty}\frac{\G [k+1]\,\G [k+2]}{\G [2k+2]}
\Bigl(\bar t\,\G[0]\Bigr)^{k+1}\Bigl(\bar t^{\dagger}\,\G[0]\Bigr)^k
\end{equation}
The result (B.10) is obtained by using Dotsenko-Fateev multiple 
2D integrals \cite{DF}. Some further transformations of the resulting 
products have been done to simplify the final expression.

A conclusion which we can draw from this calculation is that 
the multiplicative renormalizations of the touching couplings are 
given by (2.38).


\end{document}